\documentclass[a4paper,fleqn]{cas-dc}

\usepackage[numbers]{natbib}

\usepackage{graphicx}
\graphicspath{{./figs/}}
\DeclareGraphicsExtensions{.pdf,.jpeg,.jpg,.png}

\usepackage{subcaption}
\usepackage{url}
\usepackage{booktabs}

\def\tsc#1{\csdef{#1}{\textsc{\lowercase{#1}}\xspace}}
\tsc{WGM}
\tsc{QE}
\tsc{EP}
\tsc{PMS}
\tsc{BEC}
\tsc{DE}

\begin{document}
\let\WriteBookmarks\relax
\def\floatpagepagefraction{1}
\def\textpagefraction{.001}
\shorttitle{MAVIDH Score}
\shortauthors{Gomes et~al.}

\title [mode = title]{MAVIDH Score: A COVID-19 Severity Scoring using Interpretable Chest X-Ray Pathology Features}

\author[1]{Douglas P. S. Gomes}
\cormark[1]
\ead{douglas.uf@gmail.com}


\address[1]{Machine Vision and Digital Health (MAVIDH) Research group, School of Computing and Mathematics, Charles Sturt University, NSW, Australia}

\author[2]{Michael J. Horry}
\ead{michael.j.horry@student.uts.edu.au}

\author[1]{Anwaar Ulhaq}
\ead{aulhaq@csu.edu.au}

\author[1]{Manoranjan Paul}
\ead{mpaul@csu.edu.au}


\address[2]{Centre for Advanced Modelling and Geospatial Information Systems (CAMGIS), School of Information, Systems, and Modeling, Faculty of Engineering and IT, University of Technology Sydney, NSW, Australia}

\author[2]{Subrata Chakraborty}
\ead{subrata.chakraborty@uts.edu.au}

\author[3] {Manash Saha}
\ead{manash.saha@health.nsw.gov.au}

\address[3]{Wollongong Hospital, Wollongong,  NSW}

\cortext[cor1]{Corresponding author}
\fntext[fn1]{{\url{https://github.com/dougpsg/covid_mavidh_icufeatures_scoring}}}

\author[3] {Tanmoy Debnath}
\ead{tdebnath@csu.edu.au}

\author[4] {D.M. Motiur Rahaman}
\ead{drahaman@csu.edu.au}

\address[4]{National Wine and Grape Industry Centre, Charles Sturt University, Australia}


\begin{abstract}
The application of computer vision for COVID-19 diagnosis is complex and challenging, given the risks associated with patient misclassifications. Arguably, the primary value of medical imaging for COVID-19 lies rather on patient prognosis. Radiological images can guide physicians assessing the severity of the disease, and a series of images from the same patient at different stages can help to gauge disease progression. Hence, a simple method based on lung-pathology interpretable features for scoring disease severity from Chest X-rays is proposed here. As the primary contribution, this method correlates well to patient severity in different stages of disease progression with competitive results compared to other existing, more complex methods. An original data selection approach is also proposed, allowing the simple model to learn the severity-related features.
It is hypothesized that the resulting competitive performance presented here is related to the method being feature-based rather than reliant on lung involvement or opacity as others in the literature.
A second contribution comes from the validation of the results, conceptualized as the scoring of patients groups from different stages of the disease. Besides performing such validation on an independent data set, the results were also compared with other proposed scoring methods in the literature. 
The results show that there is a significant correlation between the scoring system (MAVIDH) and patient outcome, which could potentially help physicians rating and following disease progression in COVID-19 patients.
The reference code is available in our public repository \footnote{\url{https://github.com/dougpsg/covid_mavidh_icufeatures_scoring}}. 
\end{abstract}

\begin{graphicalabstract}
\includegraphics{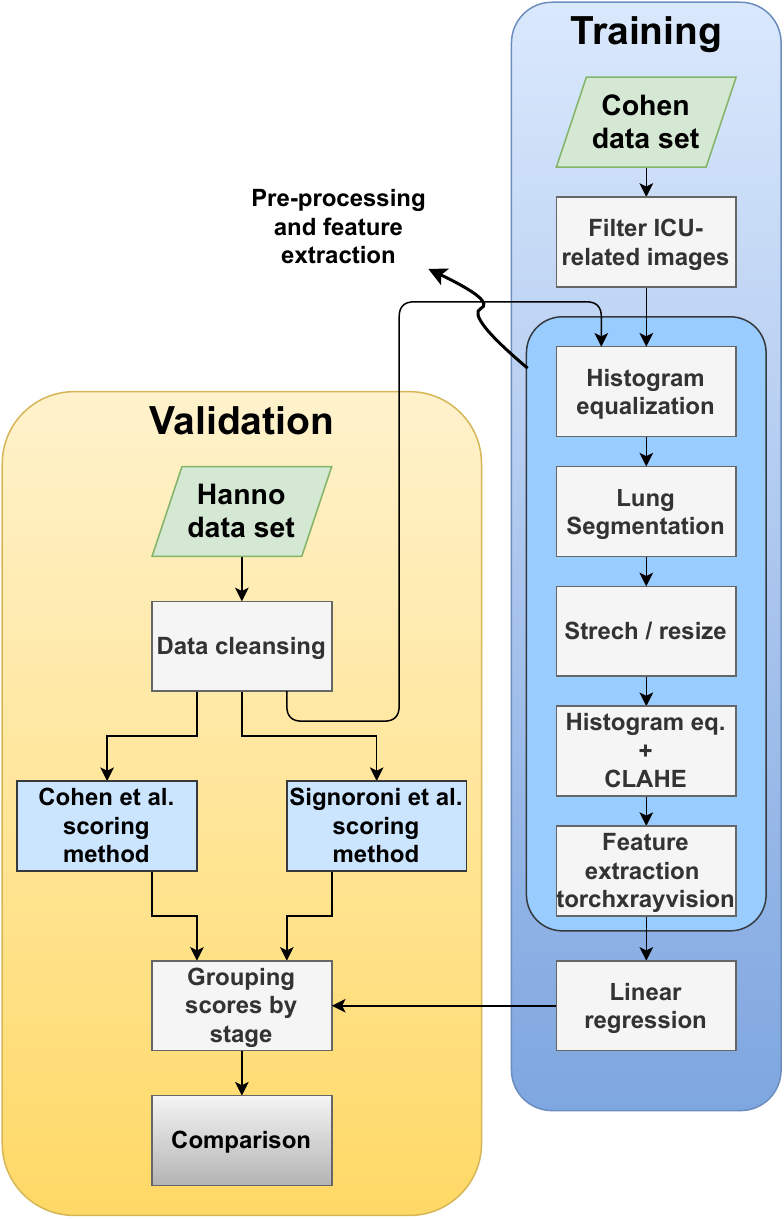}
\end{graphicalabstract}

\begin{highlights}
\item A competitive and simple method for COVID-19 scoring based on Chest X-Rays
\item Use of pathology semantic features rather the lung involvement measures
\item Methodology for validation on a data set with patients in different stages of disease
\end{highlights}


\maketitle

\sloppy

\section{Introduction}

\label{sec:introduction}
COVID-19 remains an imminent threat to society with many countries experiencing a severe second wave of the virus \cite{giacomo2020second}. One widely documented characteristic of COVID-19 infection is the wide range of symptoms experienced by infected persons ranging from entirely asymptomatic through admission to the general ward for a range of symptoms including fever, cough, fatigue, headache and diarrhoea \cite{huang2020clinical} to severe pneumonia requiring admission to ICU with mechanical ventilation \cite{guan2020clinical}. Reported case-fatality rates vary from 1\% to greater than 7\%, usually due to respiratory failure \cite{vincent2020understanding}. Fatal cases tend to progress rapidly, particularly in the case of elderly patients where the average survival time after admission can be as low as five days \cite{wang2020coronavirus}. Given this wide range of patient symptoms and potentially rapid progress to death in severe cases, it is imperative that a patient’s condition be objectively tracked for severity so that scarce critical care resources may be efficiently deployed to improve patient outcomes.

Whilst care in modern ICUs will result in death rates towards the lower end of the range, life-sustaining therapies will, in practice, be limited by lack of personnel, beds or materials and equipment. There is an emerging body of evidence showing a close association between greater mortality and overwhelmed healthcare infrastructures \cite{ho2020covid}. This limitation of resources leads clinicians to make prognostic decisions based on criterion such as old age, fragility and comorbidity that can lead to the death of patients with poor prognosis in favour of patients with better progression outlook \cite{vincent2020understanding}. Understanding patient prognosis is therefore critical to patient outcomes at both an individual and group level.

Pathological multivariate scoring has been shown to be predictive of COVID-19 patient admission to ICU and death \cite{allenbach2020multivariable, fan2020comparison}. This approach requires the collection of pathology data points such as CRP levels, lymphocyte counts, platelet counts, interleukin, and procalcitonin levels. Although effective, this approach may not be practical in a triage situation, since the collection of necessary pathological data points is both resource-intensive and time-consuming. Having a more accessible indicator for the severity progression would therefore be desirable and potentially useful.

Chest medical imaging has proven to be useful in managing more serious COVID-19 infections since progressive respiratory failure caused by massive alveolar damage is the main source of COVID-19 mortality \cite{xu2020pathological}. In particular, Chest X-rays (CXR) and Computed Tomography (CT) imaging are useful tools in the management of moderate to severe COVID-19 cases since these methods help clinicians to establish a baseline pulmonary status and identify underlying pulmonary conditions that may contribute to the patients’ risk, as well as assessing COVID-19 progression \cite{rubin2020role}. The CXR imaging mode has the advantage of being less expensive per scan in comparison to CT \cite{flores2017ma03} and available as portable apparatus that is easier to disinfect than CT equipment (which is typically fixed in a dedicated radiology room) \cite{wong2020frequency}. Although not as accurate as CT or Ultrasound \cite{horry2020}, the sensitivity of CXR imaging increases over the course of COVID-19 infection with serial CXR imaging, especially after day 6 of symptom onset \cite{stephanie2020determinants}. 

One approach to COVID-19 severity and progression scoring is the automated or semi-automated analysis of medical images taken over consecutive periods. Several such scoring techniques appear in the literature for both the CXR and CT imaging modes \cite{wasilewski2020covid, yang2020chest}. These techniques typically divide the medical image into geographical regions with each region manually assessed by radiological staff according to specific criteria. Semi-qualitative manual scoring of CXR has been shown to have prognostic value for COVID-19 progression in low, moderate/high and highly severe cases corresponding to scores 1, 3 and 4 \cite{baratella2020severity, borghesi2020covid}. Notably, there is no indication that these methods are reliable for moderate cases outside of this score range.

Since the manual interpretation of medical images is a highly specialized and resource-intensive skill, many researchers have investigated the utility of deep learning systems in determining a COVID-19 severity score. Such systems have proven to be successful in quantifying COVID-19 lung compromise at a point in time \cite{blain2020determination, cohen2020predicting}, sometimes with good correlation between results from the CXR and CT imaging modes \cite{amer2020covid}. Automation of the semi-qualitative approach has been proposed with results that are promising but leave much room for improvement, in part due to the difficultly associated with establishing the ground truth to such methods due to human radiological interpretation \cite{signoroni2020end}. The use of deep learning can have limited applications in medical use cases when used as support for decisions that affect a patient’s clinical outcome because of their non-explainable, black-box nature. One popular technique for interpretation on convolutional networks is the saliency map, which provides a heatmap overlay of network attention calculated on a gradient basis \cite{simonyan2014very}. Nevertheless, there are still serious concerns as to whether saliency mapping techniques accurately reflect trained model parameters \cite{adebayo2018sanity} since they do not have shown to be robust under rigorous examination in the context of medial imaging \cite{arun2020assessing}.

Given the issues of assessing disease progression and explainability, this work presents an interpretable, fully-automated, CXR severity scoring methodology for COVID-19 named MAVIDH (Machine Vision and Digital Health research group) score. Such a methodology uses machine learning techniques to extract semantic features from the CXR and score COVID-19 patients based on a particular selection of ICU-admission data. The validation here is performed on an independent data set and through a comparison with other existing works. The comparison uses images from patients in different stages of disease progression: not in ICU, vicinity of ICU admission, and in ICU. The analysis shows a relationship between CXR features and patient stratification. It is envisaged that this research may lead to the development of analytic tooling that will help clinicians stratify COVID-19 patients to achieve better outcomes while being valuable in assisting clinics in managing resource requirements relating to ICUs.

\section{Related work}

The amount of papers proposing computer vision techniques and other similar efforts to assist the control of the COVID-19 pandemic has established a trend in research \cite{ulhaq2020covid, manna2020covid}. Due to the popularity and accessibility of deep learning algorithms for classification tasks, one can find several works proposing algorithms for COVID-19 diagnosis based on radiological images. Nevertheless, there is a constant concern by physicians and experts that such methods have little practical use due to the potential bias, uncertainty, and consequential risks of relying on such algorithms for a diagnosis \cite{wynants2020prediction, bachtiger2020machine}. There are efforts to mitigate such problems like segmenting and aligning the lung images, which helps \cite{tartaglione2020unveiling,rabinovich2007does} but does not solve the diagnostic concerns. Some therefore argue that the most plausible fair use of such technologies is in gauging disease progressions and severity \cite{cohen2020covid}.

Regarding disease severity assessment, there have been a number of efforts to produce multivariate COVID-19 risk stratification scores \cite{wynants2020prediction}. Not all of these methods are computer-vision based, and some rely on potentially onerous data to estimate the disease severity. The access to medical imaging such as CXR and CT, represents an alternate method of COVID-19 patient risk stratification, which is to use medical images to quantify lung abnormalities. X-rays, in particular, are flexible and inexpensive technology that could help physicians assess disease progression and severity. A particularly popular approach for using CXR in disease severity is the Brixia score \cite{borghesi2020covid}, which is a method that uses a heuristic scoring system to quantify a severity score for COVID-19 pneumonia. It can be seen as a semi-qualitative assessment of lung disease by ranking pulmonary involvement over upper, middle and lower zones per lung on an 18 point severity scale. The Brixia score is an example of what will be referred to here as a lung-involvement score since it estimates to what extent the lungs are compromised by pathologies like lesions and opacities.

Works using Brixia-like scores in both automatic \cite{amer2020covid, signoroni2020end, cohen2020predicting} and non-automatic \cite{allenbach2020multivariable, baratella2020severity} approaches can be found in the literature. The intricate end-to-end methods such as \cite{amer2020covid} names their score as ‘Pneumonia Ratio’, for example; others like \cite{cohen2020covid, blain2020determination} base its severity score on features named Opacity and Geographic extension. These methods seem to have some predictive value, but most rely on the extension of lung involvement and scores given by human experts, which they try to regress. In this paper, however, a hypothesis tested is that a method that does not rely on lung involvement and expert labelled data but rather on features from known lung pathologies and data selection from patients in different stages of the disease can work well at tracking disease progression. This testing is performed by comparing the results of such a method with other existing works. However, the comparison is only possible with other methods that fit the same definition as this one: automatic, computer-vision based, using CXRs to produce the severity score, and that also had code available to be implemented in the same comparable framework. To those requirements, only two other works were found fitting such description.

Cohen et al. \cite{cohen2020predicting} employed three blinded radiologists to stage disease severity using the extent of lung involvement and degree of opacity to establish ground truth.  Then, a deep convolutional neural network, trained on a number of public datasets covering 18 common radiological findings, was used as a feature extraction layer for COVID-19 CXR images. The features were then sequentially connected to a linear regression layer to predict the extent of lung involvement and degree of opacity to labels from experts as ground truth.  The results showed relative fair measures for the regression with a mean absolute error of 1.14 for the geographic extent score and only 0.78 for the lung opacity score. Still, no validation on patients from different stages of disease progression was performed.

Taking a different approach to Cohen, Signoroni et al. \cite{signoroni2020end} used a large clinical dataset of 5000 CXR images to train a deep learning-based implementation of the semi-quantitative Brixia score. The results showed that it achieved an accuracy equivalent, or better than human radiologists with arguably greater consistency.  After segmenting and aligning lung fields from source CXR images, a preprocessing pipeline was implemented to equalize and denoise the dataset before using these images to train variations of a ResNet-18 based CNN (BS-Net). The network showed comparable performance against an independent COVID-19 dataset in portability studies indicating potential usefulness in other clinical settings.

The work proposed here has comparably slightly less complexity than \cite{signoroni2020end}, which is fully end-to-end, but it is a bit more intricate than \cite{cohen2020predicting}, which does not use lung segmentation as a preprocessing step. The few degrees of lesser complexity allows the method presented here to be more interpretable then more complex alternatives as it uses a specialized CNN to extract features but, as in \cite{cohen2020predicting}, uses such features in simpler but potentially more robust and explainable learning methods.

\section{Data and methods}

\subsection{Data sets}

The main aspects of this work are the two data sets of X-rays images from different sources and the original methodology for severity scoring based on low-complexity, semi-interpretable, machine learning. The critical data set for assessing the hypotheses presented here was recently proposed and contained a set of multiple-instances images of patients at different stages of the disease \cite{hannodata}. By having samples of images at different moments of disease progression from the same patient, this data set can be seen as a source of potential insights regarding the role of X-rays in assessing disease severity. Differently from most published in the literature up to date, this data set has rich metadata containing the distance (in days) from when the image was taken to the hospital and ICU admission. It is the most detailed data in this respect, to the best of our knowledge. Given the authors’ affiliation to Hannover Medical School in Germany, this data set will be referred here as ‘Hanno’. At no stage, this data set was used to learn the methods presented here; the Hanno data set was used only for validating the proposed hypotheses and comparing results to other methods. The other data set, used for learning the features relevant to the disease severity, is the popular set of X-ray images by Cohen et al. \cite{cohen2020covid}, referred here by the first author’s name. The Cohen data set has been used in many works in the literature focused on disease diagnosis, but much less in progression and severity scoring. The data set has limited metadata on the ICU admission in regards to the moment in time of when the image was taken but has some potentially relevant categorical classes concerning the stage and patient severity.

In short, the Hanno data set comes with 234 images, but the quality is much varied. Cleansing is performed on images where the lung field is not clearly visible, and a significant part is excluded. In total, 154 images resulted from such a selection. From these, 54 are images from patients that were not admitted in ICU, while the remaining 100 images are from patients taken in different stages of disease progression. To help dealing with size limitations, a set of gentle affine transformations was used to augment the data, pushing it to 986 images in total. This set of images is used in the same conditions for all the comparison to other methods and validation. Although it usually only contains a few images from the same patient, the rich metadata with the offset in days from key moments of disease progression can be used to group patients in defined periods of progression so an original validation approach can be adopted; more on the grouping method in the following subsections. Table \ref{tab:dataset_sum} presents the quick summary of these two data sets discussed for clarity.

The Cohen data set is a somewhat popular set of X-Ray images, first announced in April 2020. Although it was first described in March 2020, it has more than 200 citations, 2000 favourites in Github, and many implementations. A significant part of citing works use this data set to address the diagnosis problem. Nevertheless, the diagnosis of COVID through imaging is a delicate problem, and its solution should present high robustness to risk in order to be useful. A number of authors have written on the challenges of such a use \cite{arun2020assessing, degrave2020ai, wynants2020prediction} with some pointing that probably the most promising use of X-rays would be at assessing disease severity and progression in a prognostic approach \cite{cohen2020covid, manna2020covid}. With this in mind, this methodology proposes a different use of the data set, where classes of patients with different ICU-related outcomes are adopted. By using this approach, it is expected that the trained model will learn a reasonable probability estimator of patient outcome that could be used as a score. In all, after selecting the images with reasonable quality where the lungs and its features were visible and fitted our particular criteria for the patient stage, 100 images were selected. A set of the same affine transformations, applied for data augmentation as in the previous data set, pushed the number of samples to 1040 in total. The importance of the semantic features, in contrast to lung-involvement scores, were reported by the authors of this paper in a work describing how a similar approach could point to future ICU admission \cite{gomes2020potential}.

\begin{table*}
	\caption{Summary of the data sets used in fitting and validating the scoring model}
	\centering
	\begin{tabular}{lllp{3.7cm}lp{2cm}}
		\toprule
		Dataset     & Type & Location & Collection & COVID images &  \# Images after augmentation \\
		\midrule
		Cohen (fitting) & CXR  & Global & Web scraped & 468 & 1040   \\
		Hanno (validation)     & CXR & Germany &  \small Institute for Diagnostic and Interventional Radiology & 240 & 986   \\
		\bottomrule
	\end{tabular}
	\label{tab:dataset_sum}
\end{table*}

Regarding model selection, it is worth further commenting on the major factors to the methodological decisions proposed here, which is to avoid overfitting and prioritize interpretability. The main reasons for that are two-fold: (1) the data sets used are limited, and one should therefore minimize the risk of overfitting, and (2) differently from methods using end-to-end deep learning for X-ray analyses, interpretability can be much valuable in subjects like healthcare. The over-cautious care for overfitting leads to the choice of low-complexity methods (low Vapnik–Chervonenkis dimension), which are also usually more interpretable. The feature extraction role is performed by a deep, pre-trained specialized CNN, but the regression and classification results are given by a simple logistic regression model. Such a specialized CNN was trained to detect common lung pathologies with a combination of large data sets of tens of thousands of X-ray images. Its outputs are features learned to have semantic meaning, i.e., each of them can be seen as probability scores of different lung pathologies. It is thus more interpretable in the sense that the features are real pathology scores, and one can see how much weight is given to each feature in the subsequent severity scoring. This construction relates deeply with one of the hypotheses discussed here, which is that features representing specific pathologies in the lung may correlate better with the severity of symptoms than methods that estimate lung compromise, thus resulting in a more realistic scoring. Since the feature extraction method is performed by such a deep, specialized network, one can say that it does fit into the learning paradigm known in the literature as transfer learning \cite{shermin2019enhanced}.

\subsection{Feature extraction pipeline}

Before extracting the features with such specialized CNN, the selected images are fed through a preprocessing pipeline, which has the goal of normalizing and reducing potential artifacts that could lead to biases in learning. Such a pipeline comprises sequential blocks with defined tasks: histogram equalization, lung segmentation, and cropping of the lung area. All these are done automatically with processes that share the same parameters for all images. Histogram equalization is the main normalization, and it is applied at the beginning and end of the pipeline. At the input, the normalization is standard, and it equalizes the pixel distribution so they are not concentrated in a small range, which could induce bias. Before the output, an image normalization technique called Contrast Limited Adaptive Histogram Equalization (CLAHE) is applied, which could be seen as a local histogram normalization. This technique is used to highlight the small local pathology features, so their contrast could be improved.

The lung segmentation step is subsequently applied to the images in the preprocessing pipeline. As data sets are limited, and there are many bias-inducing features in the images like bones, medical devices, and letters, segmenting the lungs can be seen as increasing the signal-to-noise ratio for the learning algorithm. This practice is corroborated in previous studies \cite{rabinovich2007does} in the literature where one can also find comprehensive reviews on lung area segmentation \cite{candemir2019review}. Earlier works on lung segmentation presented dice similarity scores up to 0.989 \cite{hwang2017accurate}; a much simpler approach using the U-Net architecture was also proposed, achieving 0.974 \cite{ronneberger2015u}. The latter was trained on the JSRT dataset \cite{shiraishi2000development} consisting of 385 CXR images with gold-standard masks. In this work, we present a deep learning-based lung segmentation that surpasses such scores.

\begin{figure}[t]
  \centering
  \includegraphics[width=.8\columnwidth]{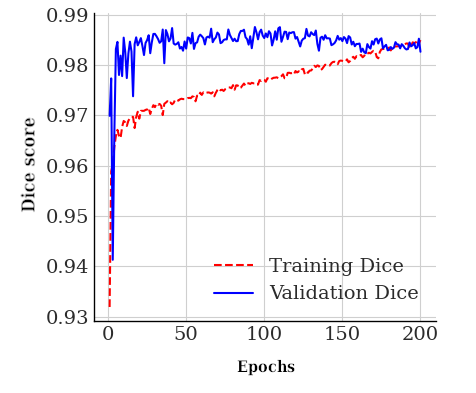}
  \caption{Training and validation curves from the lung segmentation network --- Skip U-Net.}
  \label{fig:seg_net_training_curves}
\end{figure}

The lung segmentation network learned here is a variation of the U-Net architecture using skip connections with a ResNet backbone \cite{he2016deep}. This adoption is in contrast to the U-Net trained on simple convolutional stacks with max pooling and VGG architecture \cite{simonyan2014very}. The authors hypothesized that the skip connections allow the features in earlier layers to be reused, thereby increasing the performance of the segmentation. Such a network design achieved a maximum validation dice similarity coefficient of 0.988 at epoch 93, as illustrated in Fig. \ref{fig:seg_net_training_curves}, which depicts its learning curves. The performance boost was probably also achieve given the use of additional data sets adding up to 1185 CXRs samples \cite{jaeger2014two}.

\begin{figure}[t]
  \centering
  \begin{subfigure}[t]{0.3\columnwidth}
    \includegraphics[width=\textwidth]{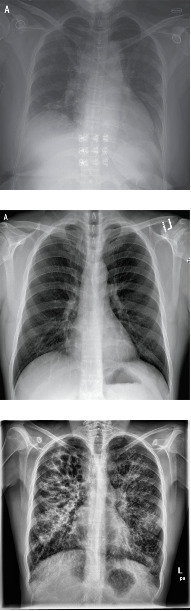}
    \caption{}
  \end{subfigure}             
  \begin{subfigure}[t]{0.3\columnwidth}
    \includegraphics[width=\textwidth]{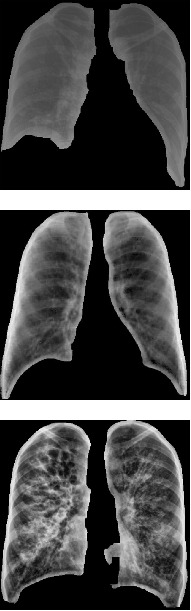}
    \caption{}
  \end{subfigure}
  \begin{subfigure}[t]{0.3\columnwidth}
    \includegraphics[width=\textwidth]{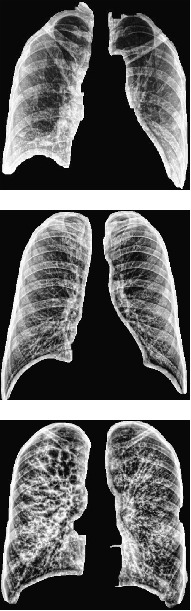}
    \caption{}
  \end{subfigure}
  \caption{Chest X-Ray images in different stages of the pre-processing pipeline. a) Original images with different contrast and histogram. b) Segmented lungs without the normalization techniques. c) Images with lungs segmented and normalization applied.}
  \label{fig:image_proc_pipeline}
\end{figure}

The pipeline then ends with lung-area cropping, which is a combination of closing of the masks and cropping any pixels outside their area. The closing of the masks helps to get rid of any artifacts escaping the lung segmentation network, and it is performed by a combination of morphological closing, contour and flood filling. The cropping has a desirable effect of normalizing the images for lung size; the lung area sometimes occupies only part of the image, either by the size the extent that the image covers of a patient or by factors like the size of the person or age. Examples of images in different stages of the preprocessing pipeline are illustrated in Fig. \ref{fig:image_proc_pipeline}.
The steps of the feature extraction pipeline can also be summarized for clarity:
\begin{itemize}
    \item Filtering ICU-related images - Selection of the images that contain ICU-outcome metadata.
    \item Histogram equalization - Equalization of pixels intensities, specially important for those with a narrow distribution.
    \item Lung segmentation - Segmentation of the lung field (see Fig. \ref{fig:image_proc_pipeline} with a specilized segmentation network (ours), which performs competitively with state-of-the-art results.
    \item Strech/resize - Operation on the lung mask so that they can fill the whole canvas image, having the effect of normalizing for different lung sizes and locations of the original canvas. 
    \item Local histogram equalization - Local equalization of the lung masks to highlight the features and patterns.
    \item Feature extraction - Use of a pre-trained specialized network that outputs semantic pathology-related scores.
\end{itemize}

\subsection{Learning and validating the progression score}

The methodology framework can be divided into two parts: learning and validation. The fact that each one of these parts is performed with data sets from different sources helps to attest the method’s ability to generalise. However, a meaningful result can only be achieved if overfitting can be limited, especially given the number of samples in the Cohen data set. Regarding the use of such a data set in the learning stage, one of the original ideas presented here is that, different from other methods that use it in a classification approach with ‘COVID’ and ‘Non-COVID’ classes, this methodology filters the metadata in particular classes: patients that were admitted to ICU but had the image taken before admission (‘future icu’) and patients that recovered without intensive care (‘not icu’). This filtering is particularly interesting because the first class is populated by samples of patients that, despite not being present in ICU at the time the image was taken, were eventually later admitted. Their images could potentially present insightful information of predecessor features to disease severity.

At the learning stage, the methodology using the semantic features for progression scoring is performed by first extracting the semantic features (specialised CNN), standardising them, and fitting a logistic regression model to classify the data in one of two mentioned classes (‘future icu’ and ‘not icu’). If the accuracy of the classifier showed it to have relevant invariance to the classes, one can hypothesise its estimated class probability (output from the logistic regression) could be correlated to the patient’s severity state. That would not be necessarily surprising since such the classifier was learned to attempt to predict if a particular patient, given their X-ray, will eventually be admitted to ICU or not. 

The original validation approach proposed here comprises two parts: (1) a heuristic method for grouping scores from patients’ images taken in similar stages of symptom progression, and (2) a comparison of the group scoring to other methods in the literature by implementing their code in the Hanno data set. It is worth noting that the grouping in the first step is only possible because, differently from Cohen (used for training), the Hanno data set contains rich metadata on the offset (number of days) from specific moments in the disease progression. The proposed heuristics here is to group X-ray images from 3 distinct periods or classes of patients:

\begin{itemize}
    \item Group 1: Images from patients not admitted to ICU.
    \item Group 2: Images from patients in the vicinity (between 1 day before or after) of ICU admission.
    \item Group 3: Images from patients currently in ICU (between 1 day past ICU admission and 1 day before ICU release).
\end{itemize}

By the definition of these groups, one can hypothesise that despite the actual value of the score, it should respect a particular trend between groups to have at least minimal significance. For example, Group 1 should have the lowest scores on average since it contemplates patients that did not go to ICU. Moreover, one should expect that Group 2, populated by images of patients in the vicinity of ICU admission, to have high scores on average as they are close to or presently presenting severe symptoms.

\begin{figure}[t]
  \centering
  \includegraphics[width=1\columnwidth]{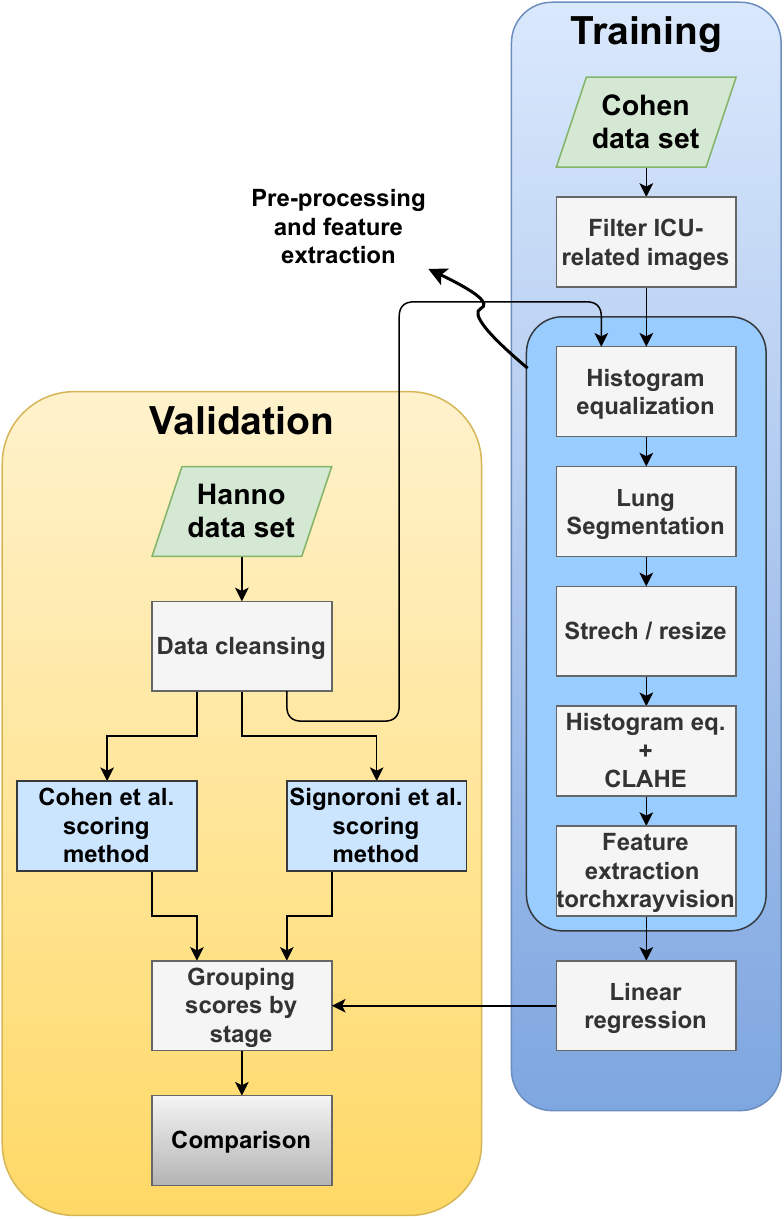}
  \caption{Graphical abstract of the methodology. Images taken from the Cohen data set are particularly selected, processed and used for learning a severity scoring model. The images from the Hanno data set are processed and scored by the learned method and by other two from the literature. The assessment of the results is performed via a grouping method also proposed here.}
  \label{fig:graph_methodology}
\end{figure}

Following the expected trend corroborates the hypothesis that it is any indicator correlated to disease severity. The contribution is reinforced if such a scoring method presents competitive results to other methods in the related literature. This point should be highlighted by noting that the data sets for training and validation come from different distributions, and any resulting validation should be more dependable than trivial train-test splits on the same data set. It is also important to note that the comparison to other methods was only possible because the authors made their code and neural network weights available. This initiative is praiseworthy and should be incentivised since it can increase the progress rate in such an urgent field. To better clarify how the pieces of the methodology work together, Fig. \ref{fig:graph_methodology} presents the synthesised methodology in diagram form.

\section{Results}

\subsection{Observational findings}
\begin{figure}[t]
  \centering
  \begin{subfigure}[t]{\columnwidth}
    \includegraphics[width=\textwidth]{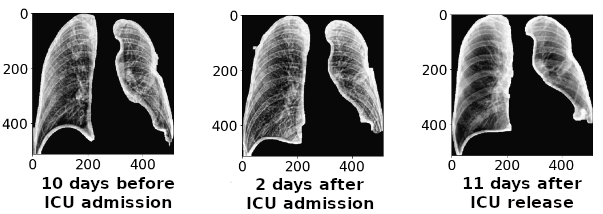}
    \caption{Example of a case where the images loosely correlate to the development of the disease. \label{subfig:obs_cases_a}}
  \end{subfigure}             
  \begin{subfigure}[t]{\columnwidth}
    \includegraphics[width=\textwidth]{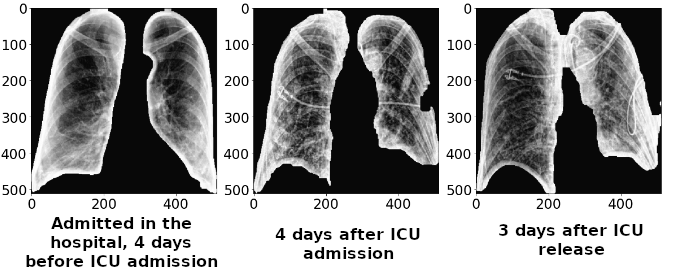}
    \caption{Example of a case where the quality of the images is counter-intuitive to the actual disease progression. \label{subfig:obs_cases_b}}
  \end{subfigure}
  \begin{subfigure}[t]{\columnwidth}
    \includegraphics[width=\textwidth]{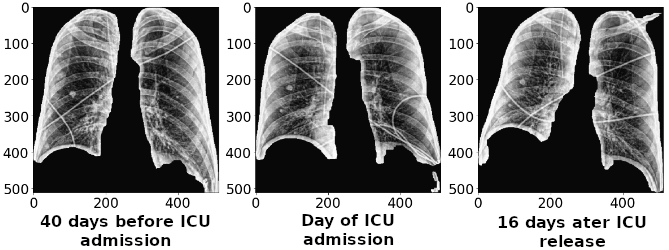}
    \caption{Example of a case where the images taken in very distinct moments do not show evident visual differences. \label{subfig:obs_cases_c}}
  \end{subfigure}
  \caption{Noticeable cases of images taken from patients in distinct moments of their disease progression.}
  \label{fig:obs_cases}
\end{figure}

The close inspection of the data sets given by the experience from developing this methodology generated insights on the potential role of X-rays to score disease severity. Although it is a somewhat trivial notion that X-rays will not translate all the information to a definite score the disease progression, some discrete insightful examples can illustrate some of the limitations; the authors have not encountered such observations in other works, especially regarding the X-ray diagnosis-related literature. These honest illustrations are not necessarily unexpected since other comorbidities and patients' conditions will significantly affect the severity of their symptoms. Some notable cases are depicted in Fig. \ref{fig:obs_cases} as examples that correspond (Fig. \ref{subfig:obs_cases_a}) or not (Fig. \ref{subfig:obs_cases_b}-\ref{subfig:obs_cases_c}) to the expected stages of the disease.

Given such findings, the question then becomes to what extent X-rays can track the severity in the disease progression. This question, as the primary motivation of this paper, is important because although other methods in the literature rate the severity of lung damage with features like opacity and geographic extension, they do not attest if such scores actually correlate to the progression of the patients' symptoms or need for intensive care. The following results from grouping the patients in similar stages and comparing their score address this issue.

\subsection{Scoring}

As a preamble to the results, it is worth noting a crucial aspect that makes this approach original, which is the data selection for learning. The scoring method itself is simple and direct, but as in every other problem, data is still paramount; model complexity is inconsequential when the labelled data does not present the information one needs. As it will be soon demonstrated, having a simpler but interpretable method also has considerable advantages. The main difference in the methodology adopted in data selection is the fact that images were not classified in ‘COVID’ vs. ‘non-COVID’, for example. They also did not have the lesions previously visually rated by humans in an attempt to model the intensity of score of lung damage. The hypothesis here is that some features may be more important than others in predicting severity than the overall area of involvement and that a specialised neural network could capture such information. The samples were therefore classified in two classes: images taken from patients previously to their ICU admission, and patients that were not admitted to ICU but were symptomatic enough to have their images taken. Both classes being patients with COVID. A classifier trained on these two binary classes will approximate the task of rating the probability that a patient will end up needing intensive care or not. 

\subsection{Separability}

Before using the selected classes as a scoring model, one should first attest if such particular classes are separable. Given the limited data set size, one should take extreme care to constrain overfitting and thus choose a simple classifier not simply to shatter the data set. Since a simpler classifier means a simpler set of admissible functions relying on fewer features, they have the advantage of being more robust with a stronger chance of generalising. 

The classifier chosen, given the model selection criteria mentioned above, was a shallow and limited decision tree. The only parameters changed from the scikit-learn library \cite{scikit-learn} defaults were the maximum depth of the tree to 4, and the minimum leaves in a node to 20. This setting aggressively limits the ability of the thee to branch into nodes based only on a few observations, consequently limiting overfitting. 
The in-depth work of using shallow decision trees for analysing the semantic features relevant for classification was done in a previous work by the authors \cite{gomes2020potential} and will thus not be discussed in detail here.

\begin{figure}[t]
  \centering
  \begin{subfigure}[t]{.8\columnwidth}
    \includegraphics[width=.8\textwidth]{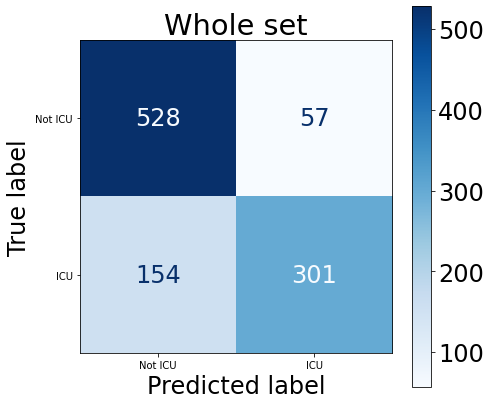}
    \caption{}
  \end{subfigure}             
  \begin{subfigure}[t]{.8\columnwidth}
    \includegraphics[width=.8\textwidth]{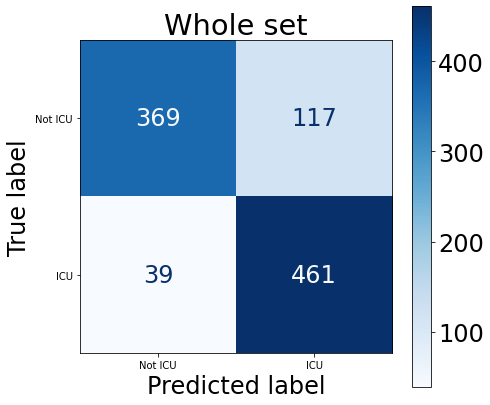}
    \caption{}
  \end{subfigure}
  \caption{Confusion matrix to assess separability in the training and evaluation data sets with the extracted semantic features: Cohen (a) and Hanno (b).}
  \label{fig:conf_mats}
\end{figure}

Fitting the shallow decision tree led to a classification accuracy of 80\% on the training data (Cohen). For cross-validation in a leave-two-out scenario, the resulting accuracy was 76\%. Note that although being from the same data set, the accuracy resulted from cross-validation does not include data used in training. It should be noted that validating on the Hanno data set it is not done in the strict sense. This data set does not have metadata on if the patient was eventually admitted in ICU in the future as does Cohen, but it has labels on which images are from patients in ICU or not. Such validation is then more related to the ability to detect severity than predicting the outcome. Having noted that, the resulting accuracy on the Hanno data set was 84\%. Both confusion matrices, from validating on the Cohen and Hanno data sets are illustrated in Fig. \ref{fig:conf_mats}. The accuracies and confusion matrices show that the data are not entirely separable, but it also indicates that there is a correlation between these 4 features ('Effusion', 'Consolidation', 'Fracture', 'Lung Opacity') and symptom severity.


\subsection{MAVIDH score assessment}

As stated in the methodology, the assessment of the MAVIDH score model was done by creating a method for grouping the patients in specific moments of symptom progression. As the score only approximately correlates with the patient severity, the grouping method helps to assess the correlation to the expected trend by comparing different statistical metrics from each group. In total, there were 54 images from patients not admitted in ICU, 27 in the vicinity of being admitted, and 72 images of patients in ICU.

\begin{figure}[t]
  \centering
  \includegraphics[width=\columnwidth]{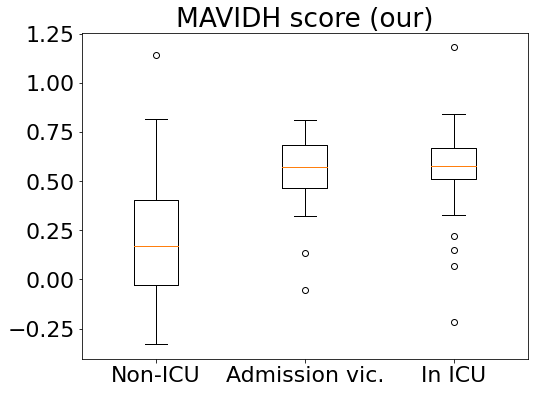}
  \caption{Group score box plot of the method proposed here (MAVIDH score).}
  \label{fig:box_score_our}
\end{figure}

Trained with the particular proposed labels, the binary regression model outputs a probability given by the linear regression function, which is used here as the MAVIDH severity score. Since every image has a respective score, the box and whisker plot was chosen to illustrate the different statistical metrics from each of the groups in Fig. \ref{fig:box_score_our}. The box plot presents some worth mentioning points regarding the agreement to the expected trend, like the fact that the upper quartile of patients that were not in ICU (0.40) was smaller than the lower quartile of all the other groups (0.47, 0.51). The fact that images of patients in the vicinity of ICU admission have more spread and almost the same median score (0.57) as the ones in ICU is also somewhat expected as these patients should be close to or already presenting severe symptoms. In the group of patients in ICU, the interquartile range consolidates at a higher level and tighter range (0.16) as compared to others, showing that severity is more consistent between images. Moreover, the caps on the ICU-related groups representing the minimum score values are all higher than the median of the non-ICU group.  

\subsection{Comparison with other methods}

The method presented by \cite{cohen2020predicting} has an interesting similarity with the one presented here, which is that both use the same feature extraction network. However, as \cite{signoroni2020end}, it tries to regress the score based on expert labelled data. In the case of \cite{cohen2020predicting}, the task is generate scores based on opacity and geographic extension labels rather than tracking patient severity. It should be noted that the following comparisons and commentary do not concern the method’s quality or ability to assess these features; they are rather related to their correlation to patient severity in the groups designed here.

\begin{figure}[t]
  \centering
  \begin{subfigure}[t]{\columnwidth}
    \includegraphics[width=\textwidth]{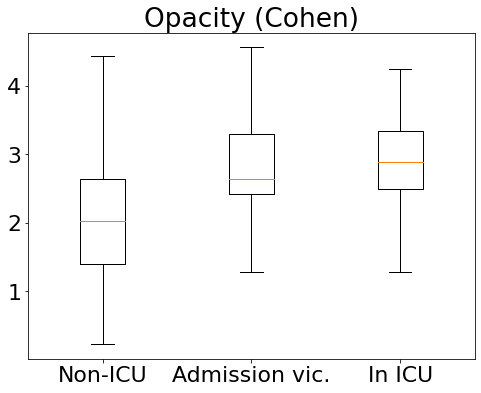}
    \caption{}
  \end{subfigure}             
  \begin{subfigure}[t]{\columnwidth}
    \includegraphics[width=\textwidth]{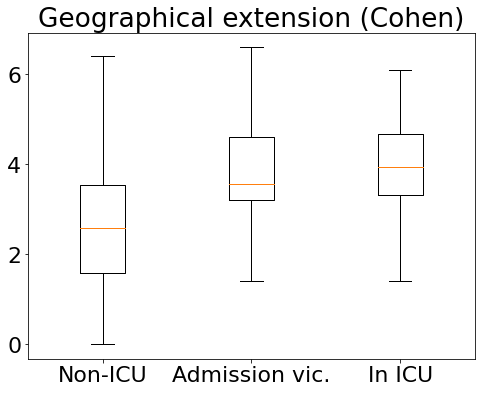}
    \caption{}
  \end{subfigure}
  \caption{Group score box plot of the a) Opacity  and b) Geographical extension features from \cite{cohen2020predicting}.}
  \label{fig:box_score_cohen}
\end{figure}

Both the opacity and geographic extension were calculated for all images in the Hanno data set using the author’s code. The opacity feature range between 0 to 6, while the geographic extention from 0 to 8. As seen in the Fig. \ref{fig:box_score_cohen}, they do not necessarily track the progression in an expected trend. Differently from the method proposed here, the range of score for ‘non-ICU’ patients are much less concentrated at a lower level and spam most of the range of other groups. 

\begin{figure}[t]
  \centering
  \includegraphics[width=\columnwidth]{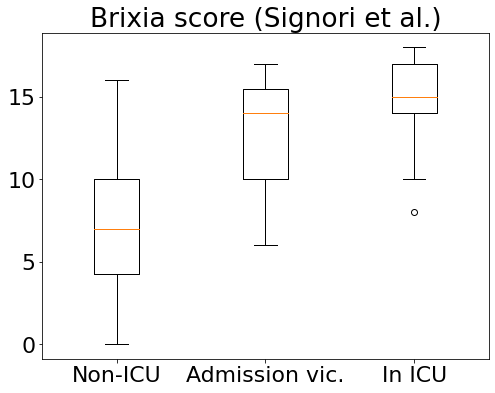}
  \caption{Group score box plot from the method presented in  \cite{signoroni2020end}.}
  \label{fig:box_score_sig}
\end{figure}

The second method to be compared presents a strong case for severity scoring. As the method being proposed here, it integrates lung segmentation but also performs alignment and rotational invariance. It is not, however, trained with severity labels but with data labelled by humans on lung compromise. The scores presented in Fig. \ref{fig:box_score_sig} were calculated by using the code and network weights kindly shared by the authors. Resulted from trying to regress scores on sections of the lung in the Brixia-score framework \cite{borghesi2020covid}, the output score ranges from 0 to 18, as it rates six different parts in scores from 0 to 3. As illustrated by the resulting box-plot scores, it is consistent at tracking the expected progression trend. Some differences from the score results from the work here can be noticed. For one, the fact that for patients not admitted to ICU, the scores are not all concentrated at a lower level with upper quartile spamming more than half the range of scores. The upper quartile is not necessarily smaller than all the other groups’ lower quartile, and the consolidation of score happens only on the group confirmed to be in ICU.  However, the scores do present a significant trend, which in the same direction as this work, attest the existence of some significance of X-ray severity scoring to help assess patient symptom severity. The authors of this scoring method should be praised since their results show some significance at tracking patient severity progression, even though that was not exactly the learning task. Nevertheless, it must be noted that this method is entirely end-to-end with interpretability reliant only on gradient maps, as opposed to the semantic feature approach in this work.

\begin{figure}[t]
  \centering
  \includegraphics[width=\columnwidth]{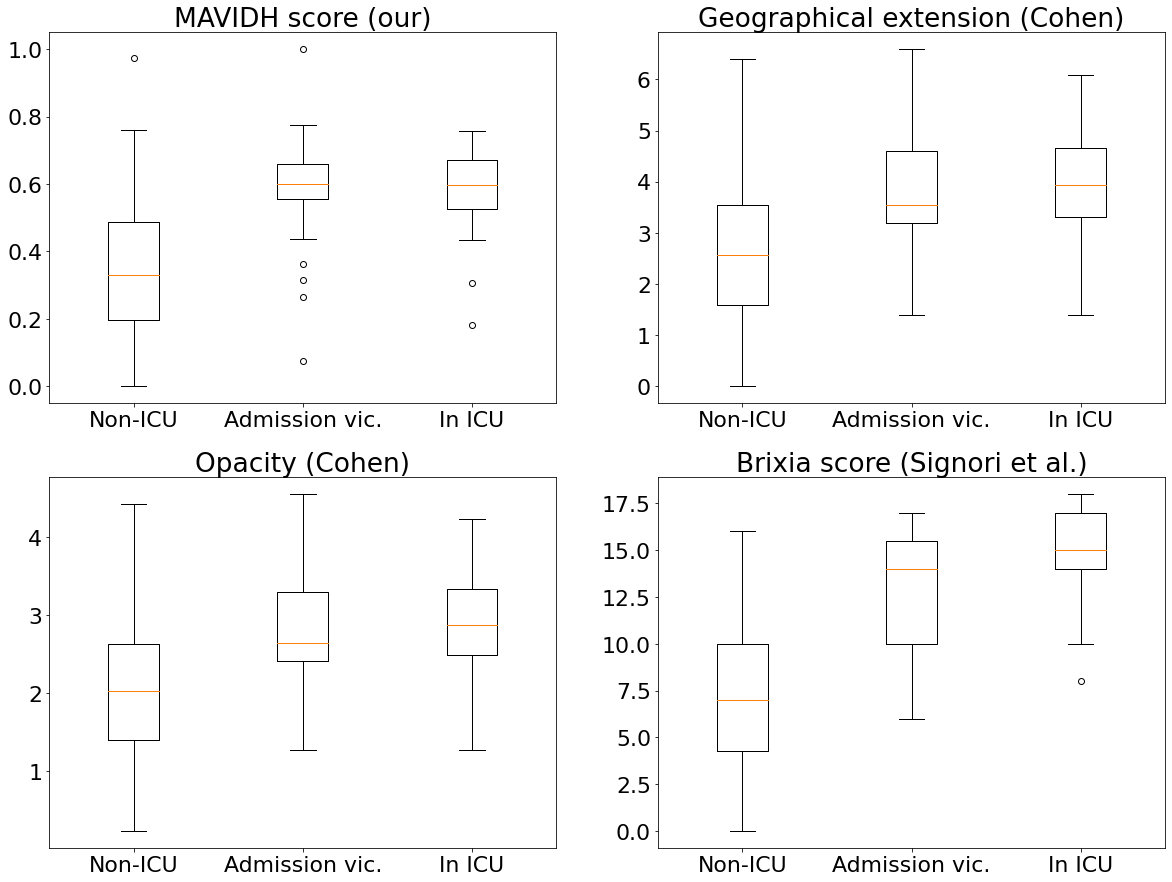}
  \caption{Comparison of the tested scoring methods.}
  \label{fig:box_score_comp}
\end{figure}

\subsection{Examples of scoring in the validation data set}

\begin{figure}[t]
  \centering
  \begin{subfigure}[t]{.85\columnwidth}
    \includegraphics[width=\textwidth]{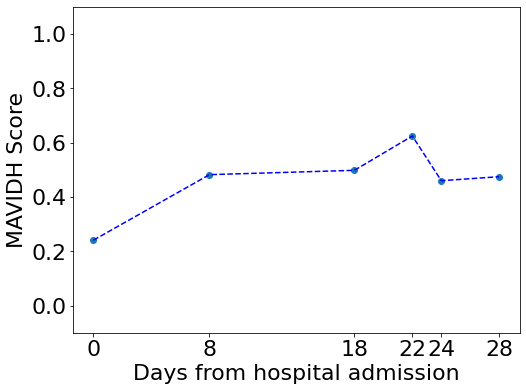}
    \caption{}
  \end{subfigure}

  \begin{subfigure}[t]{.85\columnwidth}
    \includegraphics[width=\textwidth]{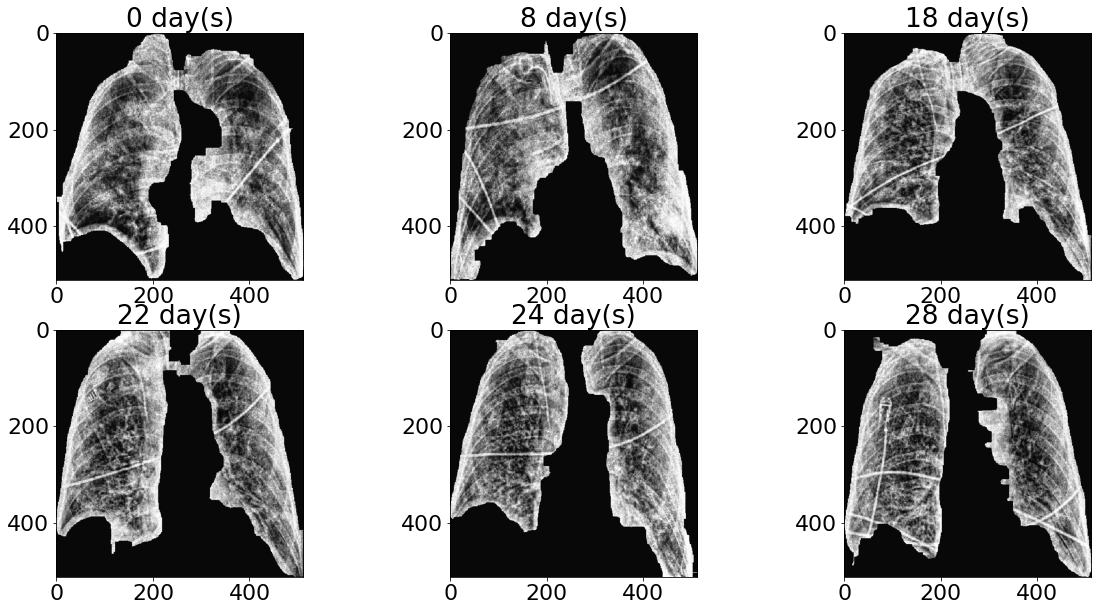}
    \caption{}
  \end{subfigure}
  \caption{Patient no. 9 scoring (a) and segmented X-rays (b). Patient still stayed in ICU for 21 days.}
  \label{fig:pat9_scores}
\end{figure}

\begin{figure}[!ht]
  \centering
  \begin{subfigure}[t]{.85\columnwidth}
    \includegraphics[width=\textwidth]{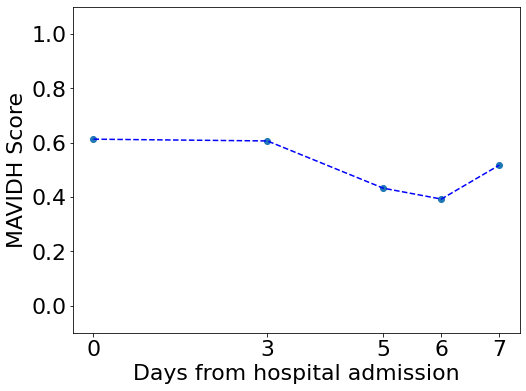}
    \caption{}
  \end{subfigure}

  \begin{subfigure}[t]{.85\columnwidth}
    \includegraphics[width=\textwidth]{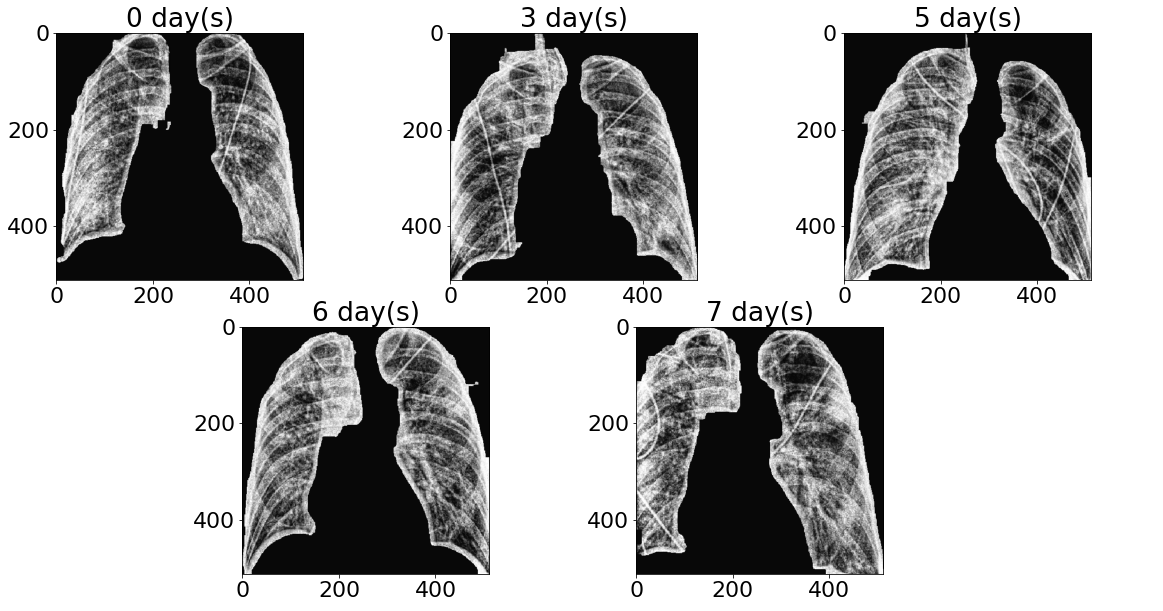}
    \caption{}
  \end{subfigure}
  \caption{Patient no. 13 scoring (a) and segmented X-rays (b).  Last image taken 7 days before ICU release.}
  \label{fig:pat13_scores}
\end{figure}

\begin{figure}[t]
  \centering
  \begin{subfigure}[t]{.85\columnwidth}
    \includegraphics[width=\textwidth]{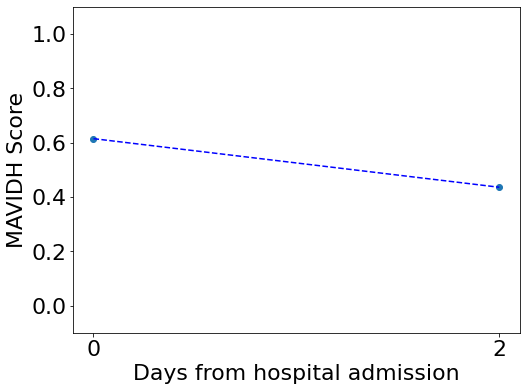}
    \caption{}
  \end{subfigure}

  \begin{subfigure}[t]{.85\columnwidth}
    \includegraphics[width=\textwidth]{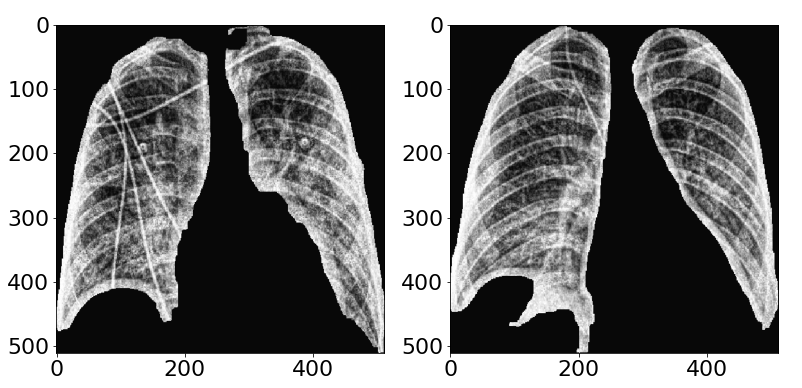}
    \caption{}
  \end{subfigure}
  \caption{Patient no. 27 scoring (a) and segmented X-rays (b). Patient was released from ICU on the next day.}
  \label{fig:pat27_scores}
\end{figure}

As a result of applying the scoring method presented here, Figs. \ref{fig:pat9_scores}-\ref{fig:pat27_scores} illustrate the scores from a particular patient through time with their segmented X-rays. The horizontal dimension of the images represent the days from the hospital admission, and a dashed line is plotted through the scores to show tendencies. Most patients had only a few images; the following illustrations are some exceptions. 

The first, illustrated in Fig. \ref{fig:pat9_scores}, was from a patient that hospitalized for 49 days. 
The last image in the plot (day 28) was taken 21 days before ICU release, and the trend shows a progressive worsening of the disease severity.
The second image, Fig. \ref{fig:pat13_scores}, shows a similar case where the patient's severity appears to fluctuate before improving. 
The patient was released 7 days after the last images were taken. 
The third image, Fig. \ref{fig:pat27_scores}, represents a simple comparison between the day the patient was admitted and two days after. 
The patient was released from ICU on the next day.


As the grouping analysis images have shown, despite the method chosen, there is probably much space for improvement in a potential COVID-19 disease severity scoring. The goal here was to show evidence that there is a probable relevant correlation in the information present in X-ray images to the severity of COVID patients rather than claiming an optimal score. It must be noted that such results were achieved with very limited data sets, and the fact that it was validated through different data sets should add to its expressivity. The authors hope to leverage better and bigger data sets to continue the investigation on the important features for disease severity that could help physicians to analyse medical imagining while assessing the patient’s progression.

\section{Conclusions}

The present work has focused on presenting a feature-based, semi-interpretable, COVID-19 disease severity scoring and comparing its significance to other methods in the literature. The scoring method comprises a feature extraction pipeline with image normalisation, lung segmentation, and feature extraction by a specialised network trained to extract semantic features related to lung pathologies. 
One notable contribution was the data selection for such learning, which is filtered and labelled in a specific way so the model can learn the severity-related information. 
Another is the fact that the method is competitive while being simplistic in its approach composed by a specialized network and linear regression on such specific data selection.
Such comparison is performed through a grouping methodology applied on a data set (Hanno) with metadata regarding the hospital and ICU admission offset. 
The analysis showed that the MAVIDH score proposed here had advantages at tracking the expected trend of the conceptualised groups. This is not to say that other methods are deficient since they were not trained for this task but rather to regress overall rate lung compromise. The results here attest the existence of a correlation between the developed score with patient severity through the disease progression. Although such a score does not perfectly correlate to severity, seen by the variance in the groups, the authors believe that it is a notable result given the limited data and the fact that presents comparable results to other more complex methods in the literature. It comprises a sensible, simple, and robust severity assessment support for a much urgent problem. The authors hope to improve this solution further with the availability of more high-quality and detailed labelled data in the near future.
More results, analyses, and source code of experiments can be found in the in work repository \footnote{\url{https://github.com/dougpsg/covid_mavidh_icufeatures_scoring}}. 

\bibliographystyle{cas-model2-names}

\bibliography{cas-refs}





\end{document}